# Estimating the Atmospheric Parameters of Early-type Stars from the Chinese Space Station Telescope (CSST) Slitless Spectra Survey

JiaRui Rao[1,2] , HaiLiang Chen[1] , JianPing Xiong[1] , LuQian Wang[1] , YanJun Guo[1,2] , JiaJia Li[1,2] , Chao Liu[2,3,4] , ZhanWen Han[1] , and XueFei Chen[1]

[1] Yunnan observatories, Chinese Academy of Sciences (CAS), Kunming, 650216, People's Republic of China; cxf@ynao.ac.cn
[2] School of Astronomy and Space Science, University of Chinese Academy of Sciences, Beijing, 100049, People's Republic of China; liuchao@bao.ac.cn
[3] Key Laboratory of Space Astronomy and Technology, National Astronomical Observatories, Chinese Academy of Sciences, Beijing, 100101, People's Republic of China
[4] Institute for Frontiers in Astronomy and Astrophysics, Beijing Normal University, Beijing, 102299, People's Republic of China
Received 2023 September 10; revised 2024 April 4; accepted 2024 April 24; published 2024 June 20

## Abstract

The measurement of atmospheric parameters is fundamental for scientific research using stellar spectra. The Chinese Space Station Telescope (CSST), scheduled to be launched in 2024, will provide researchers with hundreds of millions of slitless spectra for stars during a 10 yr survey. And machine learning has unparalleled efficiency in processing large amounts of data compared to manual processing. Here we studied the stellar parameters of early-type stars (effective temperature $T_{eff} > 15,000$ K) based on the design indicators of the CSST slitless spectrum and the machine learning algorithm, Stellar LAbel Machine. We used the Potsdam Wolf–Rayet (POWR) synthetic spectra library for cross validation. Then we tested the reliability of machine learning results by using the Next Generation Spectrum Library (NGSL) from Hubble Space Telescope observation data. In an ideal case for full-wavelength spectra without any noise, the average absolute value of the relative deviation is 2.7% (800 K) for $T_{eff}$, and 3.5% (0.11 c.g.s.) for surface gravity $\log g$. We use the spectra with the impact of interstellar extinction (AV = 0, 0.5, 1, 1.5, 2 mag) and radial velocity (RV = −50, −30, 0, 30, 50 km s$^{-1}$) from the POWR library as the test set. When RV = 0 km s$^{-1}$ and AV = 0 mag, the average value and standard deviation for 3 wavelength ranges (2550–4050 Å ($R = 287$); 4050–6300 Å ($R = 232$); 6300–10000 Å ($R = 207$)) are $-66 \pm 3351$ K, $550 \pm 3536$ K, and $356 \pm 3616$ K for $T_{eff}$, and $0.004 \pm 0.224$ c.g.s., $-0.024 \pm 0.246$ c.g.s, and $0.01 \pm 0.212$ c.g.s for $\log g$. When using the observed data from NGSL as the testing samples, the deviation of $T_{eff}$ is less than 5% (1500 K), and the deviation of $\log g$ is less than 11% (0.33 c.g.s.). In addition, we also test the influence of shifting of spectra on the parameters' accuracy. The deviation of $T_{eff}$ for the case with a shift of 5 Å and 10 Å are 3.6% (1100 K) and 4.3% (1300 K), respectively; the deviation of $\log g$ are 4.2% (0.13 c.g.s.) and 5.1% (0.15 c.g.s.). These results demonstrate that we can obtain relatively accurate stellar parameters of a population of early-type stars with the CSST slitless spectra and a machine-learning method.

*Unified Astronomy Thesaurus concepts:* Early-type stars (430); Effective temperature (449); Surface gravity (1669)

## 1. Introduction

Early-type stars, mainly including O/B-type stars (Morton & Adams 1968; Morgan & Keenan 1973; Panagia 1973), play a crucial role in various astrophysical processes. They are very important for the study of stellar and binary evolution (e.g., Heger et al. 2003; Langer 2012; Sana et al. 2012). In addition, they can evolve into many essential objects, such as black holes, neutron stars, X-ray binaries, and supernovae (e.g., Shapiro & Teukolsky 1983; Woosley & Weaver 1995; Tauris & van den Heuvel 2006; Postnov & Yungelson 2014; Han et al. 2020; Tauris & van den Heuvel 2023). Furthermore, early-type stars are important for the study of cosmic ionization and act as significant sources providing energetic feedback to the interstellar medium and intergalactic medium (Hopkins et al. 2014).

To study their formation and evolution properties, it is important to obtain their stellar atmospheric parameters, such as effective temperature $T_{eff}$, surface gravity $\log g$, metallicity [M/H], projected rotational velocity $v \sin i$, and so on

(Fitzpatrick & Garmany 1990; Langer & Maeder 1995; McErlean et al. 1999). In addition, these parameters are also important for studies on the initial mass function of young stellar population (e.g., Kroupa 2001).

The classical method of measuring stellar parameters is to compare the spectrum of the target with the model spectrum. By changing the single parameter of the model spectrum one by one, we can match the change of spectral line height, broadening, and line wing with the measured spectrum, and finally achieve the accurate measurement of stellar parameters (e.g., Trundle et al. 2007; Hunter et al. 2009; Nieva & Przybilla 2012; McEvoy et al. 2017).

Although the accuracy of traditional methods is high, it takes a long time to process the spectrum, and the processing efficiency is extremely low when facing a large amount of data. Trundle et al. (2007) investigated stellar parameters of B-type stars in the Magellanic Cloud using the high-resolution spectra from the Very Large Telescope (VLT). Hunter et al. (2009) obtained surface nitrogen abundance for 150 B-type stars in the galaxy and Small Magellanic Cloud by using the observed spectra from the VLT-FLAMES survey. Nieva & Przybilla (2012) carried out a study of the early B-type stars near the solar system by using multiple sets of high-resolution spectra. McEvoy et al. (2017) measured stellar atmospheric parameters and metallic abundance for a sample of runaway B-type stars in the galaxy to trace their formation and evolution.

Additionally, there are more and more sky survey observations that are ongoing and will be operated soon. In particular, the China Space Station Telescope (CSST; Zhan 2011) is scheduled to be launched into low earth orbit in 2024 for sky survey observation.

The CSST is a 2 m space telescope with a large field of view of ∼1.1 deg². It will simultaneously perform both photometric measurement and slitless spectroscopic surveys, which cover a large sky area of 17500 deg² in about ten years (Zhan 2011; Cao et al. 2018; Gong et al. 2019). The imaging survey has seven photometric filters (NUV, u, g, r, i, z, and y), covering from 2550 to 10000 Å, and the slitless spectrograph has three bands, GU (2550–4050 Å), GV (4050–6300 Å), and GI (6300–10000 Å), with resolution of 287, 232, and 207, respectively. As complementary to the CSST imaging survey, the CSST slitless spectra survey aims for high-quality spectra with a resolution of about 200 for hundreds of millions of stars and galaxies.

It is expected that a lot of early-type stars will be detected and their spectra will be available. This will provide an important opportunity for us to study early-type stars. Since the data from this survey is so large, it will be very inefficient to use the classical method to derive the stellar parameters of early-type stars.

On the other hand, astronomical data processing through machine learning has been blooming in recent years. Using spectral data to estimate stellar parameters is an essential application of machine learning. Machine learning has the advantages of fast processing speed and high efficiency, and is very suitable for processing massive data obtained from various large-scale sky surveys. For example, Ting et al. (2019) presents *ThePayne*, a general method for both precise and accurate estimates of stellar parameters from observed spectra, based on fitting physical spectral models.

Based on an artificial neural network, Liang et al. (2019) and Zhang et al. (2019) used APOGEE to predict the element abundance corresponding to a large number of low-resolution spectra in LAMOST survey, and cross-matched with Gaia DR2 data to determine the element abundance distribution in the region. These studies show the possibility of training models with high-resolution spectra by reducing the resolution to obtain stellar parameters with medium-resolution spectra.

In particular, Yang et al. (2020) estimated the atmospheric parameters ($T_{eff}$ and $\log g$) of DA-type white dwarfs by building a deep-learning network using the residual network. They used the three-dimensional corrected and normalized Sloan Digital Sky Survey full band spectra data (8490), of which 70% (5943) were used as training samples, and the remaining 30% (2547) were used for testing. The uncertainties of $T_{eff}$ and $\log g$ estimates are 900 K and 0.1 c.g.s., respectively. In addition, they also verified that the method applies to spectral data with resolution $R = 200$ (expected resolution of the CSST slitless spectrum), proving the feasibility of applying machine learning to CSST data processing.

To extract stellar parameters from the massive data of the LAMOST DR5, Zhang et al. (2020a) built the Stellar LAbel Machine Model (SLAM). They took the LAMOST spectral data combined with the stellar parameters from the corresponding APOGEE as training samples and obtained that the uncertainties of $T_{eff}$, $\log g$, and [Fe/H] were 50 K, 0.09 and 0.07 c.g.s., which met the demand for accurate measurement. They also selected the common targets of APOGEE and LAMOST and used the stellar parameters and spectral data of APOGEE as training samples to build a new model to test SLAM's ability to deal with high-dimensional problems, proving SLAM's ability to estimate stellar parameters by learning spectral data.

In this paper, we report our work, based on the machine learning module SLAM, to build a method to determine the stellar atmospheric parameters of early-type stars from the CSST survey spectra. This method will help researchers to quickly classify the sky survey spectral data, including but not limited to the CSST, to provide a reference for studying the physical and evolutionary characteristics of massive stars.

The paper is structured as follows. In Section 2, we describe the details of the processing of spectral data, the introduction of the SLAM model, and the method of training our SVR model by SLAM. In Section 3, we report the results of our data processing, the reliability of our model, and the results of our experiments. We will draw our conclusions and discuss them in Sections 4 and 5.

## 2. Method

In this paper, we utilized SLAM (Zhang et al. 2020a, 2020b) to measure the stellar parameters. SLAM is a forward model based on the Bayesian framework and machine learning. In SLAM, support vector regression (SVR) is employed to establish the mapping from stellar labels to stellar spectra. In other words, the inputs of this SVR model are atmospheric parameters ($T_{eff}$ and $\log g$), and the output is a corresponding spectrum. When using SLAM to estimate the atmospheric parameters for an observed spectrum, the following Gaussian likelihood is adopted

$$\ln P(\theta|f_{obs}) = -\frac{1}{2}\sum_{j=1}^{n}\left\{\frac{(f_{j,obs} - f_j(\theta))^2}{\sigma_{j,obs}^2 + \sigma_j^2(\theta)}\right.$$
$$\left. + \ln[2\pi(\sigma_{j,obs}^2 + \sigma_j^2(\theta))]\right\} \quad (1)$$

in which, $f_{j,obs}$ is the observed spectrum, $f_j(\theta)$ is the output spectrum by the SVR model for a given stellar label ($\theta$), $\sigma_{j,obs}$ is the flux error of observed spectrum, $\sigma_j(\theta)$ is the uncertainty of the output spectrum (replaced with the uncertainty of the SVR model). During this process, the trained SVR model can rapidly generate various spectra based on different atmospheric parameters. The initial values for the observed spectrum are determined by comparing it to the spectra in the training set. This comparison selects the parameters for a spectrum with maximum likelihood while considering the uncertainties of the observed spectrum. Alternatively, the initial values can also be defined based on specific scientific scenarios. Due to the high computational cost, the Markov chain Monte Carlo technique is impractical for millions of spectra. Instead, SLAM employs the Levenberg–Marquardt least-squares optimizer to compute the gradient of the likelihood function and the Jacobian matrix of the residuals. These parameters are updated iteratively, minimizing the sum of squared residuals to optimize the likelihood function. Finally, the outputs of SLAM are the stellar atmospheric parameters that maximize the likelihood function. The measurement errors of atmospheric parameters can only be roughly estimated by comparing the differences between the real values and the predicted values through statistical simulation samples (as detailed by Zhang et al. 2020a).

Moreover, SLAM has been widely applied in computing various stellar parameters with high precision. For example, Zhang et al. (2020b) used SLAM to derive $T_{eff}$, $\log g$, and [Fe/H] for K giants with low-resolution spectra from LAMOST DR5. The random uncertainties of $T_{eff}$, $\log g$, and [Fe/H] are 50 K, 0.09 dex, and 0.07 dex, respectively. Li et al. (2021) trained a SLAM model using LAMOST spectra with APOGEE DR16 labels to measure $T_{eff}$ and [M/H] for ~300,000 M dwarf stars with LAMOST low-resolution spectra. The measured $T_{eff}$ and [M/H] from SLAM, align well with the previous study determined by APOGEE observations, showing random uncertainties of ~50 K and 0.12 dex. Guo et al. (2021) utilized the TLUSTY synthetic spectral library as the training sample of SLAM and processed LAMOST medium- and low-resolution spectra. For low-resolution spectra, the measurement accuracy of $T_{eff}$, $\log g$, and $v \sin i$ is 2185 K, 0.29 dex, and 11 km s$^{-1}$, respectively; for medium-resolution spectra, the measurement accuracy of the above three parameters is 1642 K, 0.25 dex, and 42 km s$^{-1}$, respectively. Qiu et al. (2023) also trained the SLAM model to obtain $T_{eff}$, $\log g$, [M/H], and [$\alpha$/M] for M-giant stars with uncertainties of 57 K, 0.25 dex, 0.16 dex and 0.06 dex with S/N > 100, respectively. These works indicate that the synthetic library can be used as the training sample of SLAM to measure the stellar parameters of spectral data. In addition, SLAM is convenient and fast in processing large amounts of data and has high accuracy.

There are three steps in using SLAM to measure stellar parameters.

1. Data preprocessing, mainly including spectral normalization and adjustment of resolution.
2. Model training based on the spectrum with known stellar parameters.
3. Predicting stellar parameters for observed data.

### 2.1. Spectra Libraries

In this part, we mainly introduce the spectra libraries used in our work, one is the Potsdam Wolf–Rayet (POWR) spectra library based on the non-local thermodynamic equilibrium (NLTE) method (Hillier 1987; Hillier & Miller 1998, 1999), and the other is the Next Generation Spectrum Library (NGSL; Gregg et al. 2006; Heap 2008; Heap & Lindler 2010; Heap et al. 2012) from the Hubble Space Telescope Imaging Spectrograph (Woodgate et al. 1998). The 20 NGSL spectra used in this work can be obtained from the NGSL page at MAST.[5]

The training sample selected for this project is the OB model grids (Hainich et al. 2019) of POWR spectra library (Gräfener et al. 2002; Hamann & Gräfener 2003; Sander et al. 2015), which is a two-dimensional grid constructed with NLTE models. The two main parameters in the model grids are $T_{eff}$ and $\log g$. The $T_{eff}$ ranges from 15,000 to 56,000 K, and the $\log g$ ranges from 2.0 to 4.5 c.g.s. The parameter space of the POWR library is shown in Figure 1. Each grid point represents a complete synthetic spectrum with a resolution of $R = 160,000$ and corresponding stellar parameters. It can be seen that the spectral data grid points at the boundary are distributed unevenly and in small numbers, which will affect the interpolated training samples and the final machine-learning results. Therefore, it is necessary to exclude them from the interpolation grid points. After the screening, 202 grid points with their own synthetic spectra can be used for interpolation. The wavelength of the synthetic spectrum on each grid ranges from 2550 to 10000 Å, completely covering the slitless spectral wavelength range of the CSST.

We perform cubic spline interpolation between adjacent grid points to generate a three-dimensional sample space ($T_{eff}$, $\log g$ and spectral data). It can help us easily obtain the spectrum corresponding to any grid point on the two-dimensional parameter plane composed of $T_{eff}$ and $\log g$.

The advantage of using them as training samples is that a large number of high-quality spectral training samples can be obtained, as well as the corresponding precise stellar parameters, which will greatly facilitate the machine learning of spectral data with SLAM later.

The NGSL (Gregg et al. 2006; Heap & Lindler 2010; Heap et al. 2012) is selected as the observed data to estimate the future observational results of the CSST slitless spectra. The band coverage for the NGSL ranges from 1650 to 10500 Å, with a resolution of $R = 1500$. Sun et al. (2021) have proved the feasibility of reducing the resolution of NGSL spectra data to simulate the future CSST observation results on the accuracy of stellar radial velocity measurement using NGSL. Moreover, most of the target sources in NGSL are standard stars with stellar parameters, which are convenient for comparison with the measurement results as a testing sample to verify the accuracy of the model.

### 2.2. Data Preprocess

To make the model suitable for the CSST slitless spectra, we need first to reduce the resolution of the training sample to eliminate the accuracy difference. Similar to the work of Mihalas (1994), Wink & Roerdink (2004), Tay & Laugier (2008), Jennison (2013), and Sun et al. (2021), we use the Gaussian function to smooth the spectral data. We calculate the wavelength coordinate value that conforms to the logarithmic distribution under CSST-designed resolution ($R = 200$), as well as the standard deviation $\sigma$ of the Gaussian smoothing function. Then, with each wavelength point as the origin of the smoothing function, the whole spectrum is processed in the same order, and a new spectrum with the target resolution can be obtained.

When describing the full slitless spectrum of CSST, we usually take $R = 200$ as its approximate resolution. However, the resolution of each grating is slightly different because they are all tilted. Based on the private communication with the data simulation and processing team of CSST slitless spectra, 287, 232, and 207 correspond to the resolution of three bands center wavelength, which are 3370 Å of GU (2550–4100 Å), 5250 Å of GV (4000–6400 Å), and 8100 Å of GI (6200–10000 Å). In a real case, some errors may be caused by misalignment during splicing spectra. Therefore, in addition to considering the whole splice spectrum, we will also consider using the three spectra separately for our study to eliminate the influence of resolution differences of different bands. The slitless spectrum of CSST that we can obtain in the future will be assembled by the data processing group from GU, GV, and GI bands, considering the grating efficiency and the light transmission efficiency of different bands.

---

[5] https://archive.stsci.edu/prepds/stisngsl/search.php?action=Search

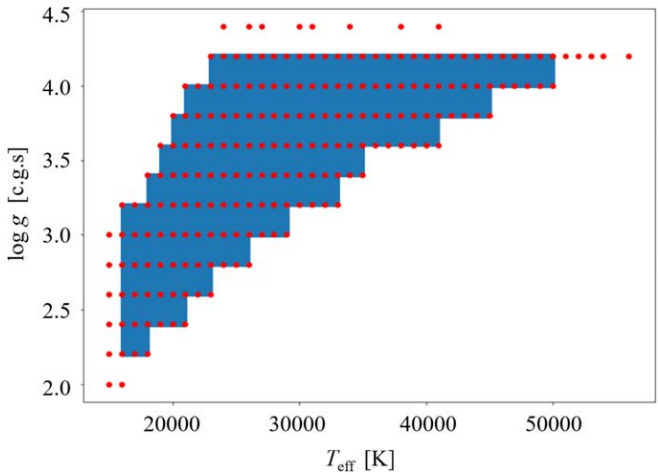

**Figure 1.** The red grid points indicate the stellar parameter space in the POWR library itself, while the blue areas indicate the parameter space for interpolation. Some grid points at the edge are not included in the interpolation range for two reasons: too few samples in the area will affect the accuracy of the model and the interpolation result is inaccurate due to missing points or uneven intervals.

**Table 1**
The Stellar Parameters of Stars in NSGL Sample

| Star Name | HD Number | $T_{\rm eff}$ (K) | log g (c.g.s) | $RV_w$ km s$^{-1}$ |
|---|---|---|---|---|
| γ Peg | 886 | 20,700 | 3.81 | 3.2 |
| ζ Cas | 3360 | 20,703 | 3.83 | −0.2 |
| AE Aur | 34078 | 31,057 | 3.54 | ⋯ |
| λ Lep | 34816 | 26,885 | 3.60 | 20.2 |
| HIP 26199 | 36960 | 27,000 | 4.10 | 27.7 |
| 15 Mon | 47839 | 32,130 | 3.47 | ⋯ |
| MCW 467 | 48279 | 31,593 | 3.51 | ⋯ |
| ε CMa | 52089 | 22,205 | 3.35 | 27.3 |
| τ CMa | 57061 | 32,514 | 3.37 | 43.0 |
| ρ Neo | 91316 | 21,576 | 3.01 | ⋯ |
| HIP 54266 | 96446 | 20,086 | 3.59 | 6.1 |
| BD+75 325 | ⋯ | 30,086 | 3.59 | ⋯ |
| θ CrB | 138749 | 16,150 | 3.75 | −25.7 |
| CD−69 1618 | ⋯ | 29,000 | 3.70 | ⋯ |
| HIP 81145 | 149382 | 27,535 | 3.92 | 3.0 |
| ι Her | 160762 | 17,789 | 3.87 | −18.9 |
| 67 Oph | 164353 | 17,574 | 2.96 | −5.2 |
| HIP 88298 | 164402 | 29,405 | 3.30 | −33 |
| HIP 92814 | 175156 | 16,361 | 3.04 | −2.8 |
| V380 Cyg | 187879 | 20,420 | 3.18 | ⋯ |

We show an example of a spectrum with reduced resolution in Figure 2, in which the gray spectrum is a high-precision spectrum with a resolution of $R = 160{,}000$, and the red spectrum is the result of segmental reducing resolution i.e., $R = 200$, 287, 232, 207 from upper to bottom, $R = 200$ corresponds to the full spectrum (2550–10000 Å), the other resolutions of 287, 232, and 207 corresponding to the center wavelength resolution of three bands. It can be seen that the main characteristics of the two spectra have the same trend. Although the operation of reducing resolution removed many details from the high-resolution spectrum, it retains strong spectral lines which play an important role in measuring stellar parameters. This result not only effectively simulates the slitless spectral observation data of the CSST, but also greatly reduces the computational complexity of the model in prediction, which is beneficial for us to process millions of survey data.

The SLAM uses normalized spectra as input since the normalized spectra regain no flux calibration while preserving the characteristics of the spectral lines. The spectra from the POWR library are normalized but those from the NGSL are the real spectra and have not been normalized. Therefore, we need to normalize the spectra in NGSL. We selected the models with parameters in the range of our training sample and removed those with strong emission lines since the strong emission lines would significantly affect the precisions of the learning result. We finally have a total of 20 spectra that could be used as the testing sample. The parameters of the 20 spectra are summarized in Table 1. In Table 1, the $T_{\rm eff}$ and log g are compiled from Koleva & Vazdekis (2012), and weighted mean absolute radial velocities (RV$_w$) are obtained from Gontcharov (2006).

Our normalization method is based on the Gaussian smoothing function (Wink & Roerdink 2004) and Cubic spline interpolation (McKinley & Levine 1998). We need to smooth the whole spectrum to eliminate the influence of emission and absorption lines and finally obtain the continuous spectrum as the denominator. Especially for certain areas where smoothing functions are not applicable (such as the Balmer jump), a small range of polynomial fitting (Bessell 1999) or even manual point selection is necessary to obtain perfect normalized spectra as much as possible. The 20 normalized spectra are presented in Appendix B (Figures 12, 13, and 14).

### 2.3. Training and Testing the SLAM Model

Before we apply the SLAM to the identified early-type stars from the CSST data to predict their stellar parameters, we first need to verify the robustness of applying the module to the training sample. The verification is achieved through the usage of a consistency check and cross-validation (CV). CV is a popular methodology used to judge the performance of machine learning models in a less optimistic approach. This approach divides the input data into $k$ groups (Ojala & Garriga 2010). Among them, the $k-1$ groups are randomly selected as the training sample, and the spectra that remained in the last group are chosen to be the testing sample, such that stellar parameters of spectra in the testing sample will be predicted by the SLAM.

In this work, we established four SLAM models corresponding to different wavelength ranges (Whole: 2550–10000 Å; GU: 2550–4050 Å; GV: 4050–6300 Å; GI: 6300–10000 Å). First, we divided the POWR library into two subsets following a ratio of 3:1 ($k = 4$). In other words, we randomly extract 25% of the 50,000 spectra (i.e., 12500 spectra) obtained after reducing resolution and interpolation as the testing group. And the other 75% (i.e., 37,500 spectra) of the data will be used as the training sample. These spectra are trimmed according to the corresponding wavelength ranges. The hyperparameters C, $\epsilon$, and $\gamma$ are set as [10, 100], 0.05, and [0.1, 0.01], respectively. Finally, the testing sample is used to evaluate the performance of our SLAM models.

The synthetic spectrum is an idealized model with a very high signal-to-noise ratio (S/N) and cannot simulate the influence of various factors in actual observations. There may be a slight shift in the central wavelength of the spectrum during the wavelength calibration for observed spectra, and reddening may affect the measurement of parameters. Additionally, the effects of extinction and radial velocity vary at different wavelengths. Below, we considered how extinction and radial velocity (RV) shift affect the measurement of parameters in GU, GV, and GI bands. The CSST survey primarily focuses on high Galactic latitudes (Zhan 2021) where the extinction in the $V$-band ranges from 0 to 2 mag, with a medium value around 0.5 mag (Larson & Whittet 2005). The precision of radial velocity measurements for OB stars is 10 km s$^{-1}$ at S/N = 100, which is obtained by using the empirical spectra of NGSL to simulate the CSST stellar spectra at $R = 250$ (Sun et al. 2021). This precision may be lower than 10 km s$^{-1}$ for the observed data. To better match the observations, the larger RV dispersion should be used in the tests. Thus, in our simulation, the extinction ranges from 0 to 2 mag, and the RV shift ranges from −50 to 50 km s$^{-1}$. Then, we followed the following strategy to apply extinction and RV shift to the test set.

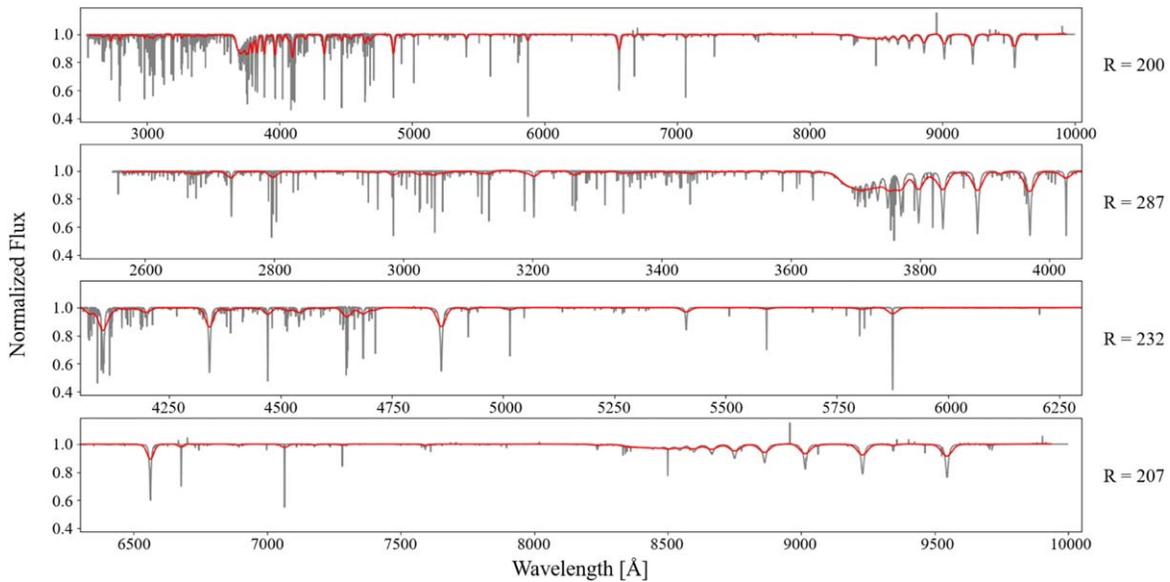

**Figure 2.** Comparison of a high-precision spectrum with the spectrum with reduced resolution. The gray spectrum is a high-precision spectrum with a resolution of $R = 160{,}000$ ($T_{\rm eff} = 34{,}000$ K, $\log g = 4.0$ c.g.s), and the red spectrum is the result of reducing resolution ($R = 200, 287, 232, 207$).

We selected the extinction values at 0, 0.5, 1, 1.5, and 2 mag, so we first randomly divided the test set (i.e., 12500 spectra) into 5 groups (i.e., 2500 spectra). Subsequently, we applied the five extinction values to the spectra of five different groups, respectively. During incorporating extinction, we followed the methodology from Cardelli et al. (1989) to convert the flux change of SED into S/N on the normalized spectrum (the S/N of the spectrum obtained from the idealized test set is assumed as 100), then added corresponding errors to the spectrum based on S/N. Then, we have selected five specific values (−50, −30, 0, 30, and 50 km s$^{-1}$) in the range of −50–50km s$^{-1}$. Following the application of extinctions, we further divided the spectra with the same extinction value (i.e., 2500 spectra) into five parts (i.e., 500 spectra) equally and randomly. Then, we applied RV shifts to the test sets. Finally, we repeat the above steps on three bands. Based on these simulated data, we can assess the influence of extinction and RV on the accuracy of predicted parameters. The evaluation of each condition was determined by the mean deviation and standard deviation between the predictions and true values ($prediction - true$). In Figure 4, we have compared the RV shifts' effects with two extinction scenarios (0 and 1 mag). We find that the deviations are relatively small for RV ranging from −50 to 50 km s$^{-1}$ except in the case with larger extinction ($Av >= 1.0$ and GU band). This means that the impact of RV shift on model accuracy is less significant compared to the impact of extinction. In the Appendix A, Table 2 summarizes the performance of model predictions with various extinctions and RV shifts. In Table 2, each data in the table consists of two parts: the average deviation between the predicted results and the true values of the test set, and their standard deviations. And we use "∼" to replace the points in the table where the standard deviation is more than 5000 K (for $T_{\rm eff}$) or 0.5 c.g.s (for $\log g$). Due to a significant decrease in the S/N of spectra in the GU band as extinction increases, the results in the GU band are expected to be poorer compared to the results in the other two bands.

Figures 3(a) and (b) illustrate the results' variations of $T_{\rm eff}$ and $\log g$ due to extinction for these spectra with RV = 0. Here, we did not show the results for other values of RV, since the results are similar. In Figure 3, the blue squares, black triangles, and red circles depict the distributions of the differences between predictions and true values ($prediction - true$) under different extinction scenarios for GU, GV, and GI, respectively. The error bars are the standard deviations of $prediction - true$. In the plots, some points have very large values and are out of the plot. Since these points have very large errors, our model is not suitable for these spectra. It can be seen that the mean deviations between predictions and true values of $T_{\rm eff}$ and $\log g$ are increased with the increasing extinction values. Compared to GV and GI, the accuracy becomes much worse in the GU band, due to the significantly dropped S/N in the GU band, specifically, when $A_V = 1$ mag, the S/Ns of GU, GV, and GI are 30, 55, and 80, respectively. Due to the minimal impact of extinction on the red arm, even with an extinction value of 2 mag, we still can roughly conduct a parameter estimation using the GI band, with the mean and standard deviation of $310 \pm 7489$ K and $-0.122 \pm 0.461$ c.g.s for $T_{\rm eff}$ and $\log g$.

Figure 4 illustrates the results' variation with RV shifts and extinctions. In Figure 4, the two columns are for two different values of extinction. Here, we did not show the results for other values of extinction since they are similar to those in these plots. From these plots, we can find that the absolute errors are relatively small for RV ranging from −50 to 50 km s$^{-1}$ except the case with larger extinction ($AV >= 1.0$) and GU band. This means that the impact of RV shift on model accuracy is less significant compared to the impact of extinction. Due to a significant decrease in the S/N of spectra in the GU band as extinction increases, the results in the GU band are expected to be poorer compared to the results in the other two bands.

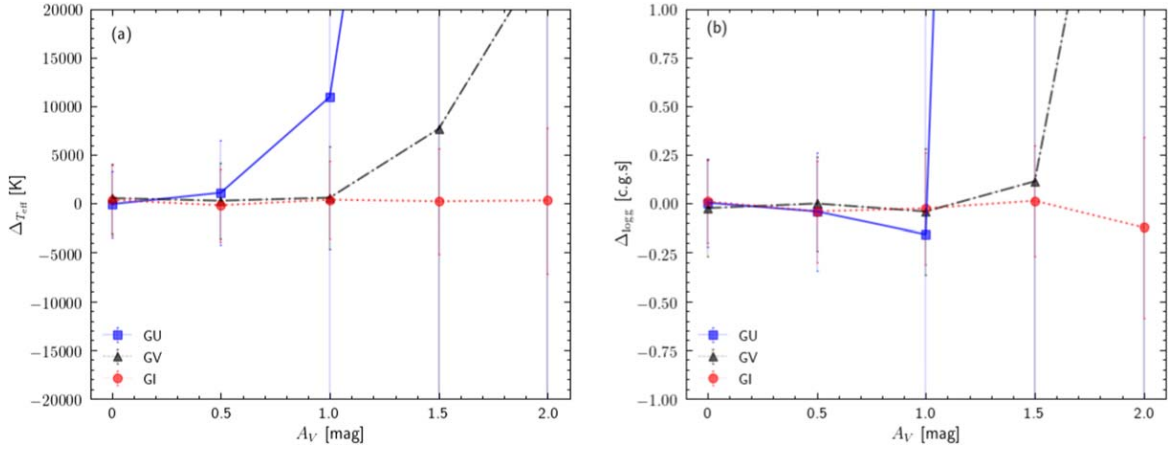

**Figure 3.** The results' variations in $T_{\rm eff}$ (panel (a)) and $\log g$ (panel (b)) with extinctions. The blue squares, black triangles, and red circles depict the distributions of the differences between predictions and true values (*prediction − true*) under different extinction scenarios for GU, GV, and GI, respectively. The error bars are the standard deviations of *prediction − true*.

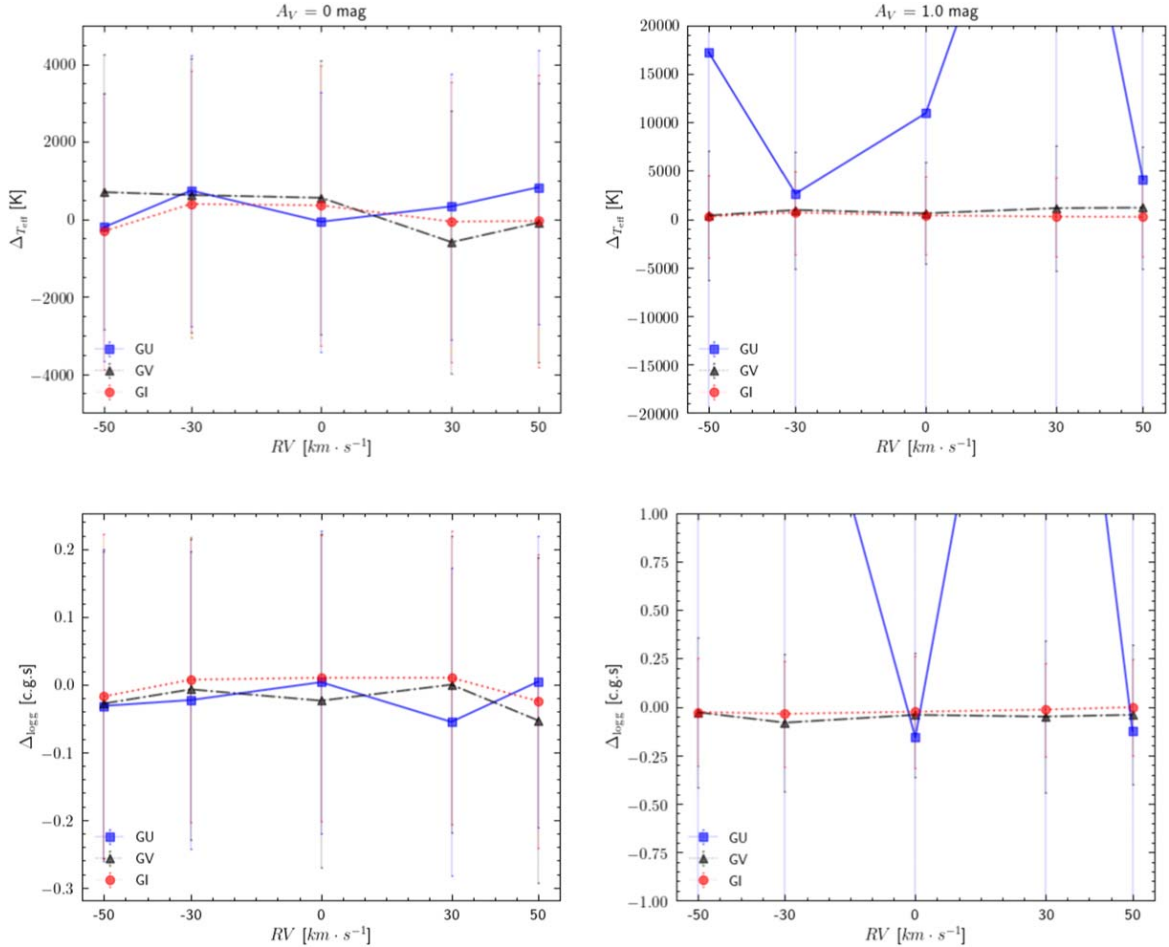

**Figure 4.** The results' variations with RV shifts and extinctions. The two columns represent different extinctions, with each panel depicting the differences between predicted results and true values with RV shifts under different extinction conditions. The blue squares, black triangles, and red circles depict the distributions of the differences between predictions and true values (*prediction − true*) for GU, GV, and GI, respectively. The error bars are the standard deviations of *prediction − true*.

Furthermore, Figure 5 shows the difference in spectra after adding Gaussian noise with different S/Ns to the same spectrum. It can be seen that when the S/N varies from 60 to 20, the entire spectrum is gradually distorted. After testing on samples with a low S/N, the absolute value of measurement deviations of the parameters ($T_{\mathrm{eff}}$ and $\log g$) we obtained are larger than 50%. Therefore, we set the lower limit of the S/N to 60.

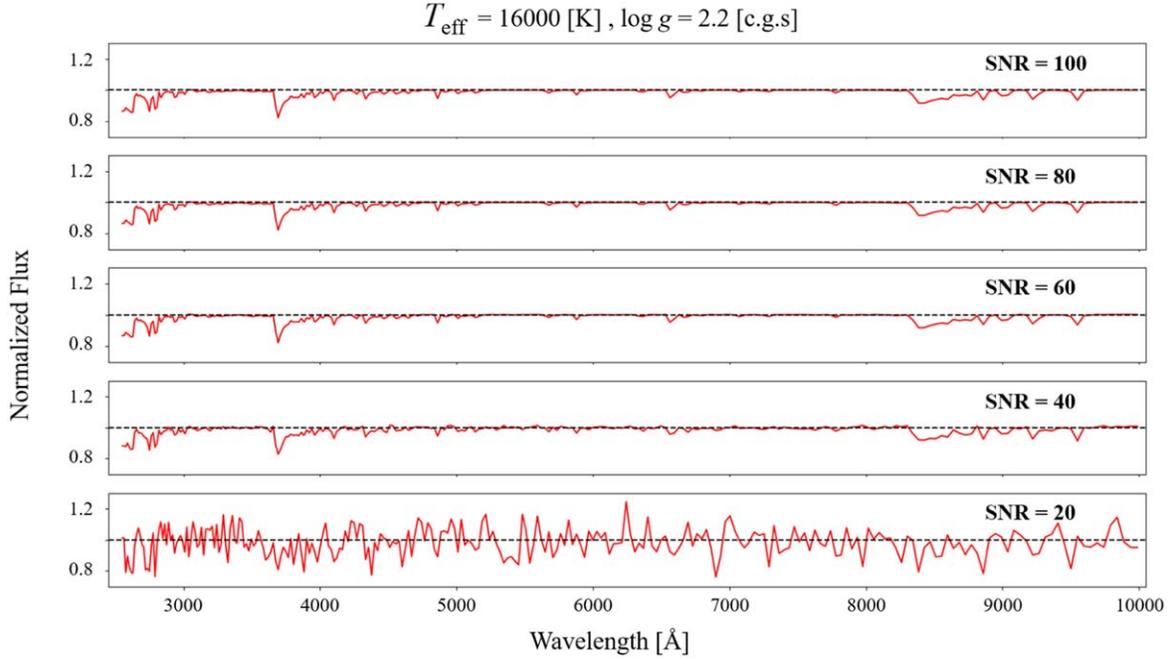

**Figure 5.** Comparison of spectra with Gaussian noise of different S/Ns added. The stellar parameters for this spectra are $T_{\mathrm{eff}} = 16{,}000$ K and $\log g = 2.2$ c.g.s.

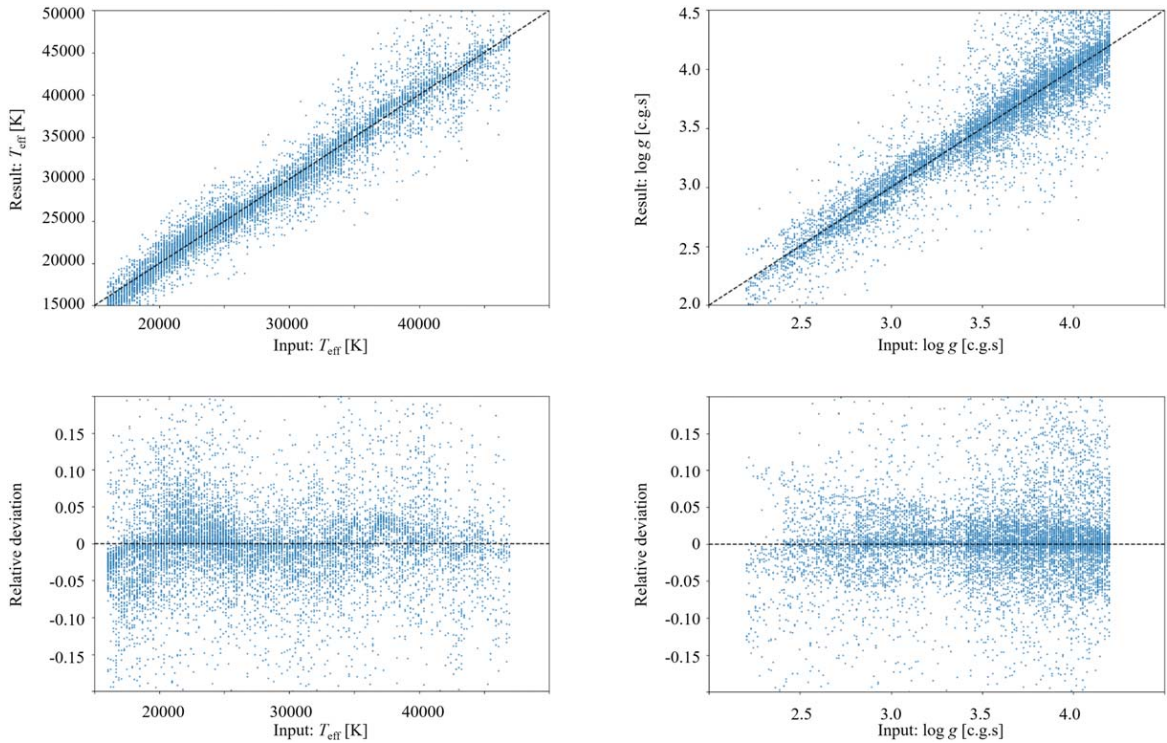

**Figure 6.** Comparison of stellar parameters of predicted values of $T_{\mathrm{eff}}$ and $\log g$ of different spectra ($y$-axis) and input parameters ($x$-axis). The upper two panels show the average relative deviations. And the lower two panels are for the relative deviation distribution histogram of $T_{\mathrm{eff}}$ and $\log g$.

## 3. Result

We use four-fold CV to perform a self-consistency check on the model. In Figures 6 and 7, we show the relative deviation distribution of the average predicted parameters of the model and the true parameters of the test samples after different treatments on the sample library.

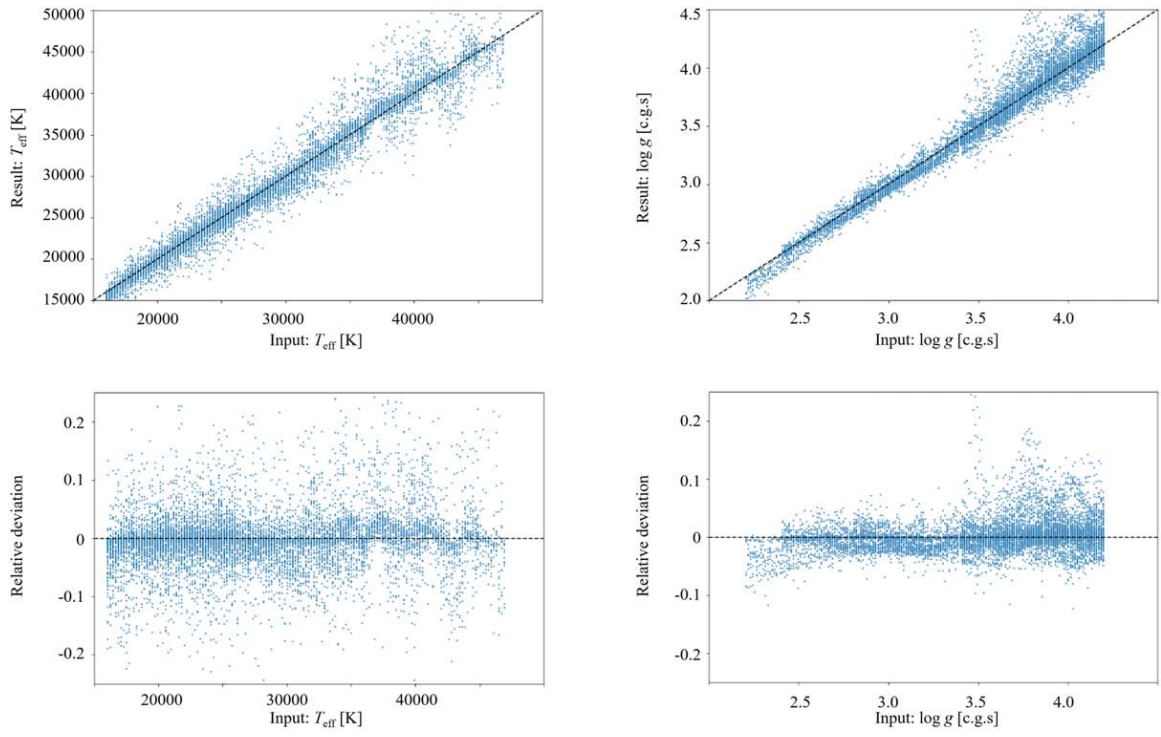

(a)

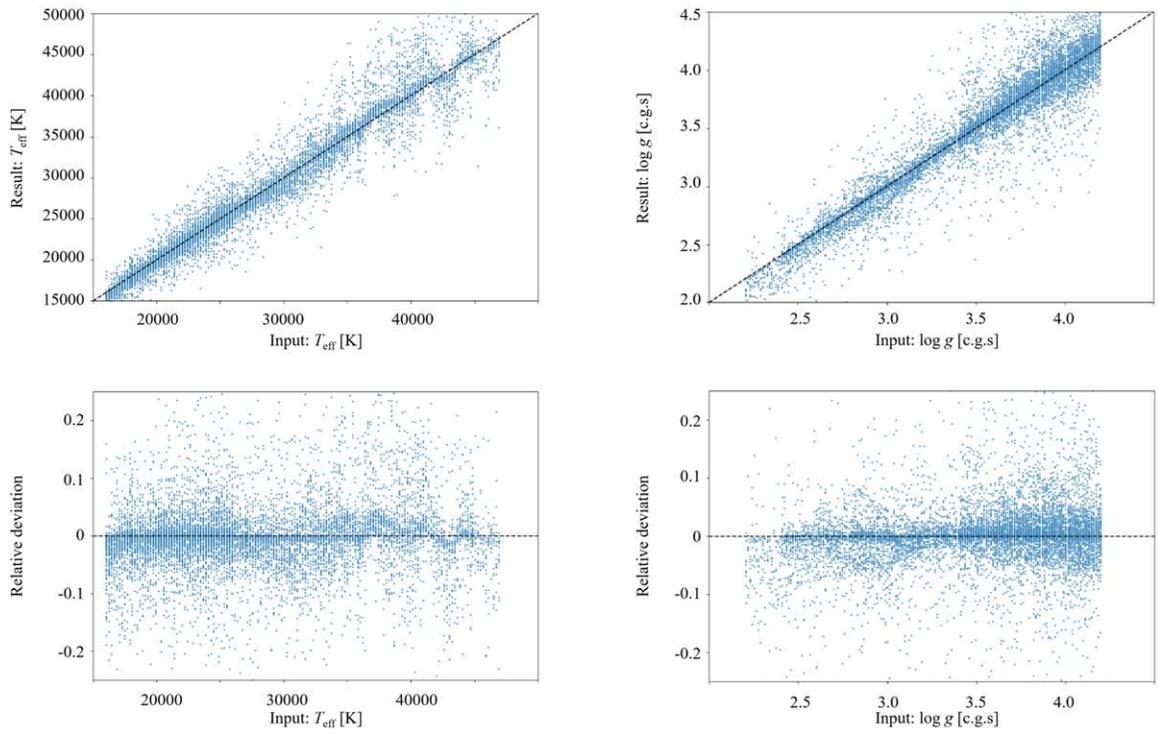

(b)

**Figure 7.** Similar to Figure 6. But the panel (a) is for the case with a shift of 5 Å, and the panel (b) is for the case with a shift of 10 Å.

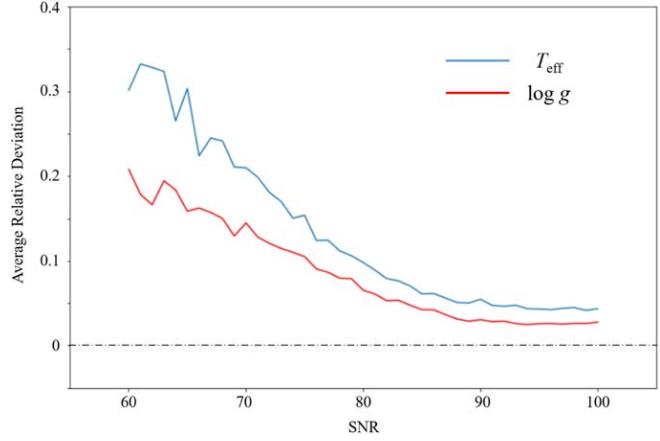

**Figure 8.** Dependence of average relative deviations of $T_{\rm eff}$ and log $g$ on the S/N. The blue and red lines are for $T_{\rm eff}$ and log $g$, respectively.

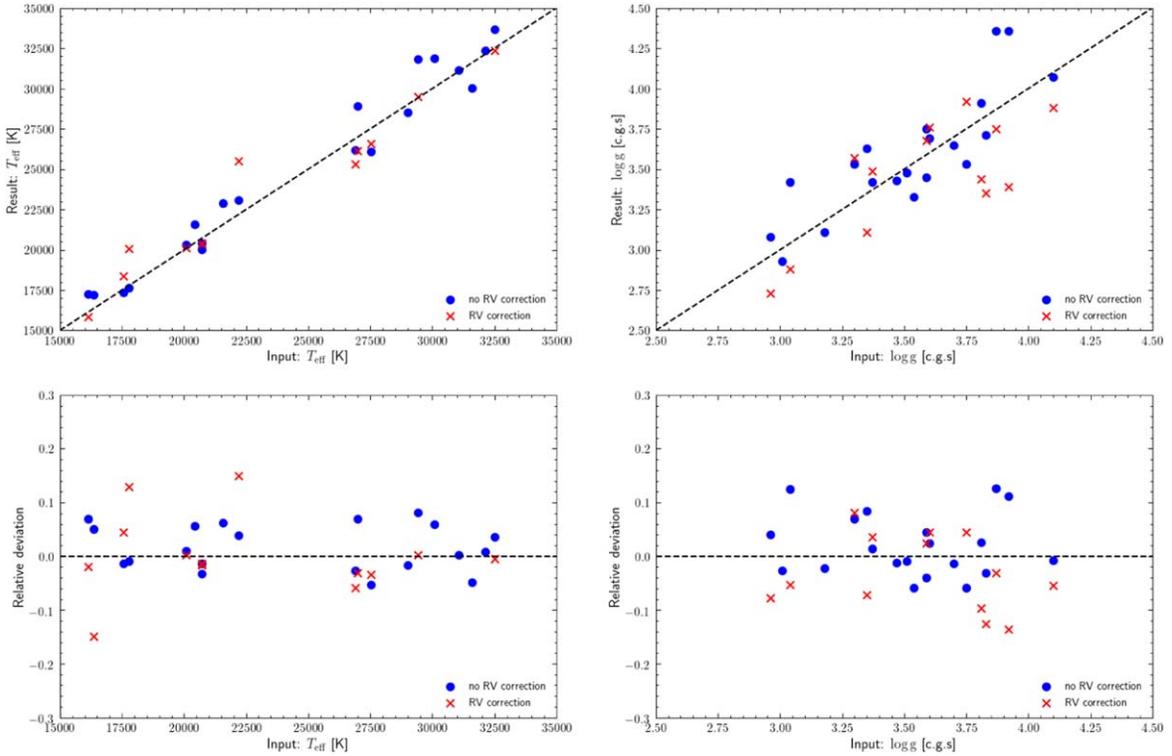

**Figure 9.** Comparison of stellar parameters of NGSL spectra obtained with our trained SLAM model (y-axis) and classical method from Koleva & Vazdekis (2012; x-axis). The lower panels show the relative deviations.

First, Figure 6 shows the ideal case where we perform cross-validation using reduced resolution synthetic spectra ($R = 200$) over the full wavelength range without adding noise. The average value of the absolute relative deviation of $T_{\rm eff}$ is 2.7% (800 K). The same value of log $g$ is 3.5% (0.11 c.g.s).

After dividing the spectra into three groups according to wavelength range, we performed four-fold CV on each group to perform a consistency check. Within one group, we average the results of cross-validations gave us to obtain two distribution histograms ($T_{\rm eff}$ and log $g$).

On the other hand, during the process of calibrating the observation results, there may be a small shift in the central wavelength of the spectrum. To understand its influence, we shift the full spectra of the testing sample by 5 Å or 10 Å. We chose to test the overall shift of the spectrum by 5 and 10 Å because, in the actual slitless spectrum observation process, it is usually difficult for us to find the position of the diffraction zero-order image point. It means that there may be a deviation of about one pixel in the position of the wavelength zero point. On the main focal plane of the CSST survey module, one pixel corresponds to a wavelength deviation of about 7 Å. Therefore, we chose to test the impact of slitless spectrum shifts of 5 and 10 Å on the final test results.

Then we cross-validate the shifted spectra and obtain a relative deviation distribution in Figure 7, which is similar to Figure 6. We can find that the average relative deviation of $T_{\rm eff}$ is 3.6% (1100 K), and of log $g$ is 4.2% (0.13 c.g.s) in the case with a 5 Å shift; the average relative deviation of $T_{\rm eff}$ is 4.3% (1300 K), and of log $g$ is 5.1% (0.15 c.g.s) in the case with a 10 Å shift.

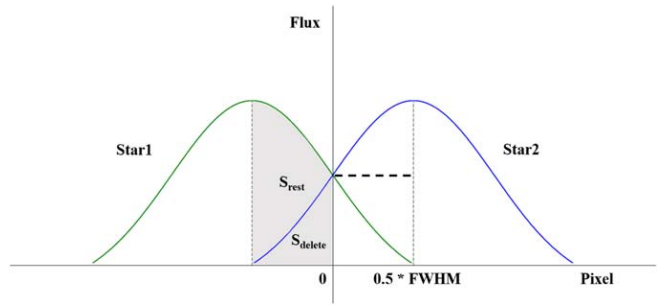

**Figure 10.** The schematic diagram of blending for two near targets. The green and blue curves represent the energy distributions of the slitless spectrum of two mixed target sources perpendicular to the dispersion direction.

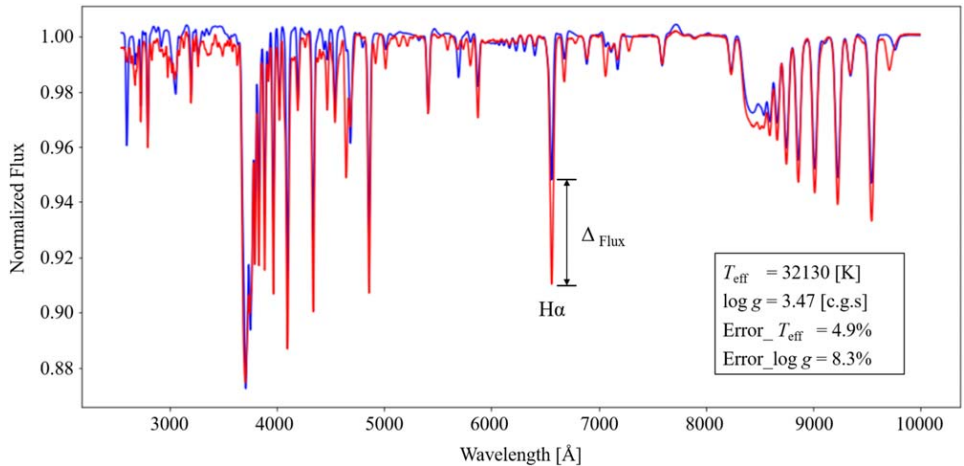

**Figure 11.** The spectra with the most significant deviation of the predicted $T_{eff}$ (4.9%, 1500 K) and the second largest deviation of the predicted $\log g$ (8.3%, 0.25 c.g. s). The blue line indicates the original spectra ($T_{eff} = 32{,}130$ K, $\log g = 3.47$ c.g.s) from the POWR library and the red line is for the spectra predicted from our trained SLAM model. At the H$\alpha$ line, normalized $\Delta$ Flux has the biggest deviation of 4%.

Moreover, we studied the average absolute value of the relative deviation of every spectrum with the same S/N (between 60 and 100) in training samples and testing samples, for assessing model ability and reliability. The results are shown in Figure 8. From this Figure, we can find that the average relative deviation decreases as the S/N increases, which is expected. This indicates that the trained SLAM model performs well and can be used for measuring the stellar parameters for real data.

In addition, we use the NGSL spectra to assess the performance of the trained SLAM model. Koleva & Vazdekis (2012) applied a full spectrum fitting approach to characterize the NGSL spectra and inferred the stellar parameters with the classical method. The precision of the parameters of the O/B stars in their results is 1000 K and 0.24 c.g.s for $T_{eff}$ and $\log g$, respectively. Before measuring the $T_{eff}$ and $\log g$ for these NGSL spectra, we need to shift the spectra to RV = 0 km s$^{-1}$. According to the radial velocities obtained from Gontcharov (2006), we corrected the RV of these spectra to 0 km s$^{-1}$. However, we only found radial velocities for 13 of these targets in the catalog. Therefore, we also compared the measured parameters derived from our model without correcting the radial velocities with the values reported in Koleva & Vazdekis (2012). In Figure 9, we compared the predicted results from our trained SLAM model with the results from Koleva & Vazdekis (2012). The blue circles represent the results derived from the spectra without RV correction, and the red crosses are the results derived from the spectra with RV correction. It can be seen that most of the relative deviations are less than 15% for $T_{eff}$ and $\log g$. When compared with these cases with and without RV correction, we can see the differences are relatively small. This is mainly because the RVs of the stars in the NGSL sample are relatively small. It is worth noting that the radial velocity can be obtained from the CSST spectra, and we can shift the spectra to RV = 0 km s$^{-1}$ when we apply our SLAM model to measure the stellar parameters.

The result proves that the trained SLAM model is suitable for spectra with a resolution of around 200 to derive stellar parameters. It meets the needs of future practical applications and can achieve the scientific goal of preliminary classification of the CSST slitless spectra survey data based on stellar parameters.

## 4. Discussion

In a slitless spectral survey, interference from other stars near the target source may be encountered. For the mid to high latitudes of the CSST sky survey, in most cases, there are not too many detectable stars and galaxies in the slitless spectra per square arc minutes in the sky. Meanwhile, since our test targets are mainly OB-type stars, their number in the Universe is relatively small. Therefore, blending will only occur when the star field is very dense. Sometimes, their impact can be removed by using software from the data processing team. The slitless spectrum has a normal distribution symmetrical about the central axis at each wavelength pixel. Its maximum value depends on the flux at that wavelength. We assumed that two-spectrum blending occurs in the GV band (with RV $= 0 \, \text{km s}^{-1}$, AV $= 0$ mag) to simulate this scenario. As shown in Figure 10, the green curve on the left represents the energy distribution (standard normal distribution) of the slitless spectrum of one of the two mixed target sources (Star1, O-type) perpendicular to the dispersion direction. When the brightness difference between the two sources is too large, the brighter one will mask the dark one, which will not have a big impact on the accuracy of our measurements. Therefore, we assume that the blue curve on the right represents another star participating in the blending (Star2), which has the same spectral energy distribution as Star1. And the distance between their maximum flux points is exactly one FWHM. For Star 1, including Star 2's components will cause an increase in the flux within the overlap region of Star 1. Based on comparing the original simulated flux and the flux after adding Star 2, we can calculate the change of S/N after this blending portion and modify the S/N in this segment. We simulated 500 blending spectra and used our model to predict stellar parameters. The results indicate that in the presence of blending, the mean deviations between the predicted values and the true values for $T_{\text{eff}}$ and $\log g$ are $672 \pm 4107$ K and $-0.031 \pm 0.371$ c.g.s. Without blending, the mean deviations of $T_{\text{eff}}$ and $\log g$ are $550 \pm 3539$ K and $-0.024 \pm 0.246$ c.g.s. We believe that the spectral data loss is acceptable.

As for how to filter out OB-type stars from hundreds of millions of survey target sources, we will rely more on the classification software being developed by the CSST front-end data processing department. However, classification through software usually mixes in about 10% of A-type stars. When their influence is reflected in the model prediction results, it mainly appears as discrete points with large deviations near the lower limit of 15,000 K. Meanwhile, the SLAM model itself has a certain extrapolation ability and can screen the mixed A-type stars to a certain extent.

In general, atmospheric parameters are closely related to the shape and depth of absorption lines from normalized stellar spectra. For early-type stars mainly, the neutral helium and hydrogen lines can be used to estimate effective temperature, while the hydrogen lines are sensitive to surface gravity (Dufton et al. 1999). We checked the spectra with the most significant deviation of the predicted $T_{\text{eff}}$ (4.9%, 1500 K) and the second largest deviation of the predicted $\log g$ (8.3%, 0.25 c.g.s.). The result is shown in Figure 11 in which the most significant deviation of normalized $\Delta$ Flux at the H$\alpha$ line (6563 Å) is marked. The difference of the Paschen series (from $\lambda = 8210$ Å) also has an essential influence on measurement results of $T_{\text{eff}}$ and $\log g$.

There is no good way to handle spectra with solid emission lines in the SLAM model. Therefore, the training samples and testing samples we used in this project are all screened. It will inevitably limit the application of our parameter measurement model. In future research, we expect to adopt more reasonable models to put the spectra with strong emission lines into the processing range.

## 5. Conclusion

Early-type stars are of great importance in many aspects. It is expected that many early-type stars will be detected in the CSST survey. It is necessary to derive their stellar atmospheric parameters based on the spectra. In this work, we discuss whether we can achieve this with a machine-learning method. By making use of the spectrum library of early-type stars, POWR, we train the SLAM model and evaluate its performance with spectra from the POWR library and NGSL of the Hubble Space Telescope (HST). The main conclusions are as follows.

1. We find that the average value and standard deviation of $T_{\text{eff}}$ derived from synthetic spectra in three wavelength ranges (2550–4050 Å; 4050–6300 Å; 6300–10000 Å) are $-66 \pm 3351$ K, $550 \pm 3536$ K, and $356 \pm 3616$ K, respectively. The values of $\log g$ are $0.004 \pm 0.224$ c.g.s, $-0.024 \pm 0.246$ c.g.s, and $0.01 \pm 0.212$ c.g.s if we use the POWR library to test the trained SLAM model when RV $= 0 \, \text{km s}^{-1}$ and AV $= 0$ mag.

2. In addition, we use the NGSL of HST with a reduced resolution $R = 200$ to test the trained SLAM model and find that the relative deviation of $T_{\text{eff}}$ and $\log g$ are less than 5% (1500 K) and 11% (0.33 c.g.s), respectively.

3. We also study the influence of a small shift in the spectra of the testing sample. We find that the average relative deviation in the case with a shift of 5 Å of $T_{\text{eff}}$ and $\log g$ are 3.6% (1100 K) and 4.2% (0.13 c.g.s.), respectively. And in the case with a shift of 10 Å, the average relative deviation of $T_{\text{eff}}$ and $\log g$ are 4.3% (1300 K) and 5.1% (0.15 c.g.s.), respectively.

4. We conclude that we can use our trained SLAM model to obtain the stellar parameters of the population of early-type stars from the CSST survey in the future. With these stellar parameters, we can further study the properties of early-type stars.

## Acknowledgments

We thank the anonymous referees for comments that helped improve the manuscript. This work is supported by the National Natural Science Foundation of China (grants Nos. 12125303, 12288102, and 12090040/3, 12073071), the National Key R&D Program of China (grant No. 2021YFA1600403), the Natural Science Foundation of Yunnan Province (grants Nos. 202201BC070003, 202101AW070003), the International Centre of Supernovae, Yunnan Key Laboratory (No. 202302AN360001), the CAS light of West China Program, the Yunnan Revitalization Talent Support Program—Science & Technology Champion Project (grant NO. 202305AB350003) and Young Talent Project. This work is also supported by the China Manned Space Project of No. CMS-CSST-2021-A10. L.W. acknowledges the support by the NSFC under program No. of 12103085.

## Appendix A
## Model Prediction Result

Table 2 summarizes the performance of model predictions with various extinctions and RV shifts. In Table 2, each data in the table consists of two parts: the average deviation between the predicted results and the true values of the test set, and their standard deviations. We use "∼" to replace the points in the table where the standard deviation is more than 5000 K (for $T_{\text{eff}}$) or 0.5 c.g.s. (for logg).

**Table 2**
Summarizing for the Performance of Model Predictions with Various Extinctions and RV Shift

| Parameters | AV(mag) | RV(km s$^{-1}$) | GU(2550–4050 Å) | GV(4050–6030 Å) | GI(6030–10,000 Å) |
|---|---|---|---|---|---|
| $T_{\rm eff}$ (K) | 0 | −50 | −206 ± 3451 | 696 ± 3554 | −308 ± 3558 |
| | | −30 | 739 ± 3500 | 622 ± 3536 | 391 ± 3437 |
| | | 0 | −66 ± 3351 | 550 ± 3539 | 356 ± 3616 |
| | | 30 | 328 ± 3435 | −594 ± 3400 | −64 ± 3616 |
| | | 50 | 823 ± 3526 | −90 ± 3596 | −45 ± 3773 |
| | 0.5 | −50 | ∼ | 213 ± 4130 | 198 ± 4168 |
| | | −30 | ∼ | 470 ± 3982 | −237 ± 3796 |
| | | 0 | ∼ | 284 ± 3854 | −181 ± 3666 |
| | | 30 | ∼ | 382 ± 4047 | 180 ± 3881 |
| | | 50 | 529 ± 4192 | −217 ± 4047 | 64 ± 4057 |
| | 1.0 | −50 | ∼ | ∼ | 280 ± 4218 |
| | | −30 | ∼ | ∼ | 680 ± 4287 |
| | | 0 | ∼ | ∼ | 388 ± 3994 |
| | | 30 | ∼ | ∼ | 262 ± 4067 |
| | | 50 | ∼ | ∼ | 237 ± 4083 |
| | 1.5 | −50 | ∼ | ∼ | 565 ± 4361 |
| | | −30 | ∼ | ∼ | 351 ± 4511 |
| | | 0 | ∼ | ∼ | ∼ |
| | | 30 | ∼ | ∼ | −10 ± 4815 |
| | | 50 | ∼ | ∼ | ∼ |
| | 2.0 | −50 | ∼ | ∼ | ∼ |
| | | −30 | ∼ | ∼ | ∼ |
| | | 0 | ∼ | ∼ | ∼ |
| | | 30 | ∼ | ∼ | ∼ |
| | | 50 | ∼ | ∼ | ∼ |
| logg (c.g.s) | 0 | −50 | −0.032 ± 0.229 | −0.028 ± 0.228 | −0.017 ± 0.24 |
| | | −30 | −0.023 ± 0.22 | −0.007 ± 0.222 | 0.007 ± 0.21 |
| | | 0 | 0.004 ± 0.224 | −0.024 ± 0.246 | 0.01 ± 0.212 |
| | | 30 | −0.055 ± 0.228 | 0 ± 0.219 | 0.01 ± 0.217 |
| | | 50 | 0.004 ± 0.216 | −0.053 ± 0.24 | −0.024 ± 0.217 |

**Table 2**
(Continued)

| Parameters | AV(mag) | RV(km s$^{-1}$) | GU(2550–4050 Å) | GV(4050–6030 Å) | GI(6030–10,000 Å) |
|---|---|---|---|---|---|
| | 0.5 | −50 | −0.018 ± 0.276 | 0.035 ± 0.26 | 0.032 ± 0.248 |
| | | −30 | −0.022 ± 0.262 | −0.025 ± 0.268 | −0.001 ± 0.232 |
| | | 0 | −0.041 ± 0.301 | 0 ± 0.243 | −0.04 ± 0.26 |
| | | 30 | −0.039 ± 0.271 | −0.022 ± 0.263 | 0.002 ± 0.244 |
| | | 50 | −0.016 ± 0.238 | −0.001 ± 0.277 | −0.003 ± 0.25 |
| | 1.0 | −50 | ∼ | −0.028 ± 0.388 | −0.026 ± 0.276 |
| | | −30 | ∼ | −0.081 ± 0.357 | −0.036 ± 0.272 |
| | | 0 | ∼ | −0.041 ± 0.321 | −0.025 ± 0.288 |
| | | 30 | ∼ | −0.05 ± 0.393 | −0.014 ± 0.241 |
| | | 50 | ∼ | −0.041 ± 0.361 | −0.001 ± 0.248 |
| | 1.5 | −50 | ∼ | ∼ | −0.027 ± 0.281 |
| | | −30 | ∼ | ∼ | −0.016 ± 0.24 |
| | | 0 | ∼ | ∼ | 0.014 ± 0.285 |
| | | 30 | ∼ | ∼ | 0.028 ± 0.281 |
| | | 50 | ∼ | ∼ | −0.037 ± 0.291 |
| | 2.0 | −50 | ∼ | ∼ | ∼ |
| | | −30 | ∼ | ∼ | −0.057 ± 0.439 |
| | | 0 | ∼ | ∼ | ∼ |
| | | 30 | ∼ | ∼ | ∼ |
| | | 50 | ∼ | ∼ | ∼ |

**Note.** (mean value±standard deviation).

## Appendix B
## NGSL Spectra

The 20 normalized spectra are presented in Figures 12, 13, and 14.

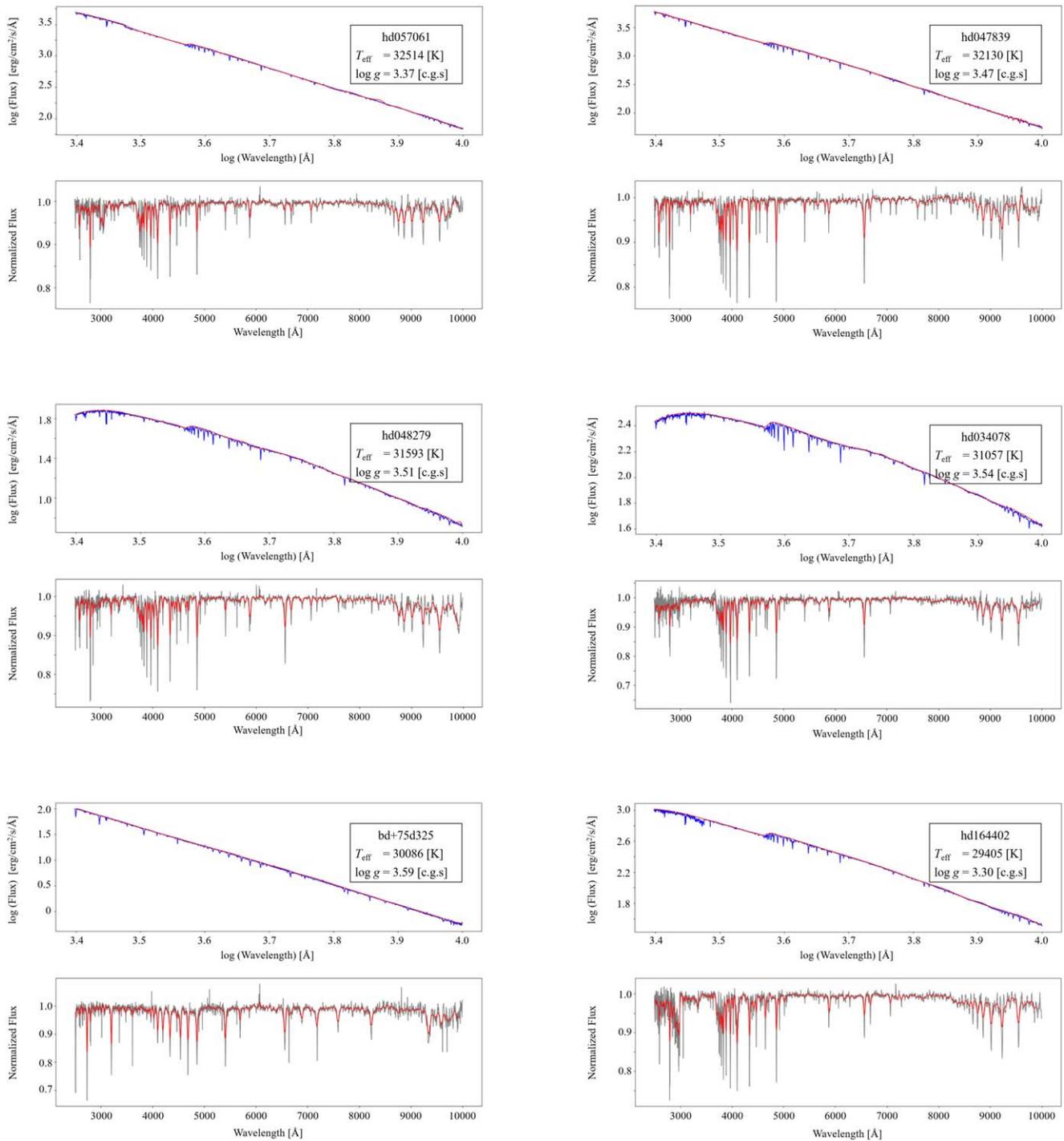

**Figure 12.** The upper panel is the spectrum in NGSL (blue), and its primitive function for normalization (red). The bottom panel is the result of normalization with $R = 1500$ (gray) and $R = 200$ (red).

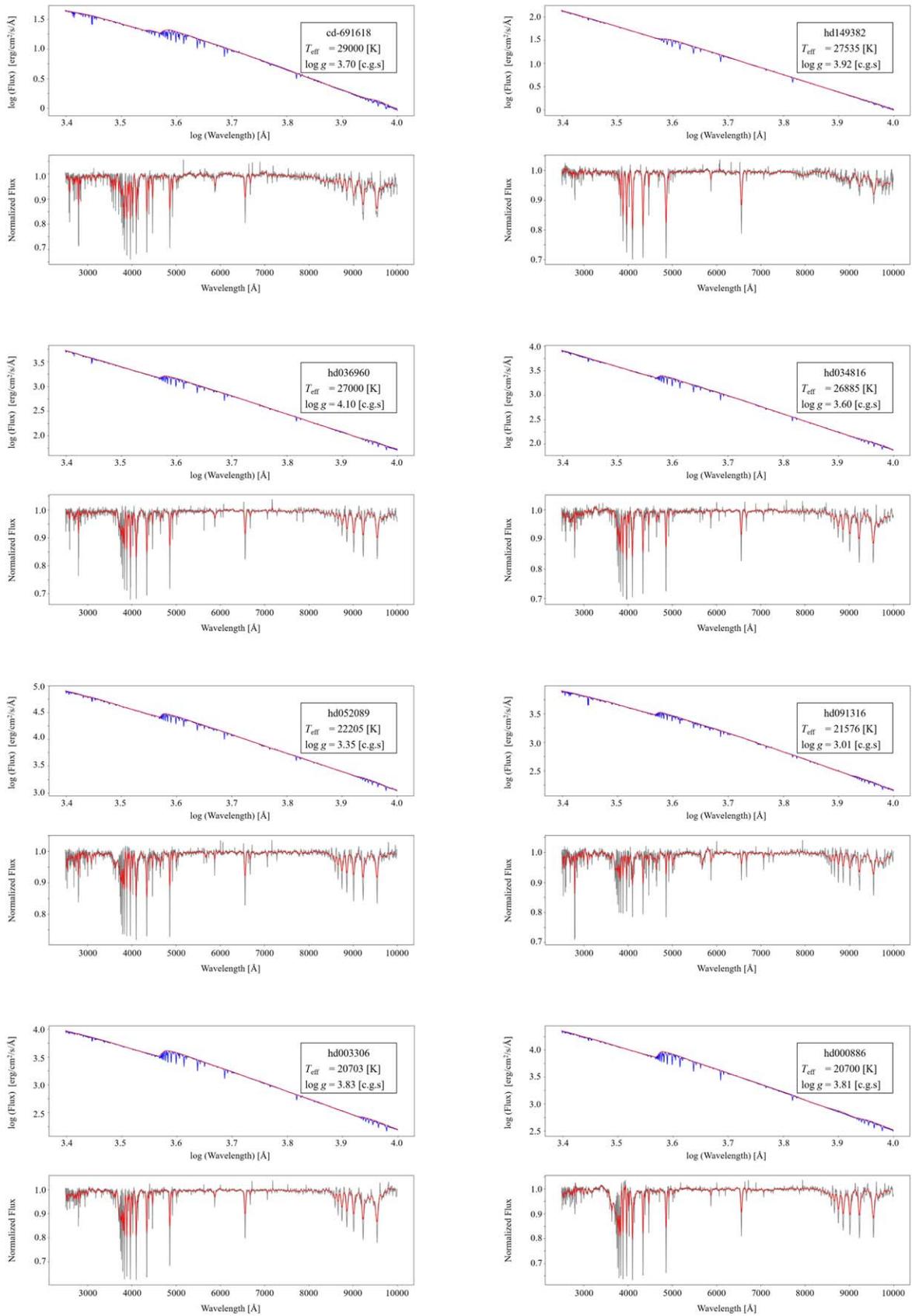

**Figure 13.** Similar to Figure 12.

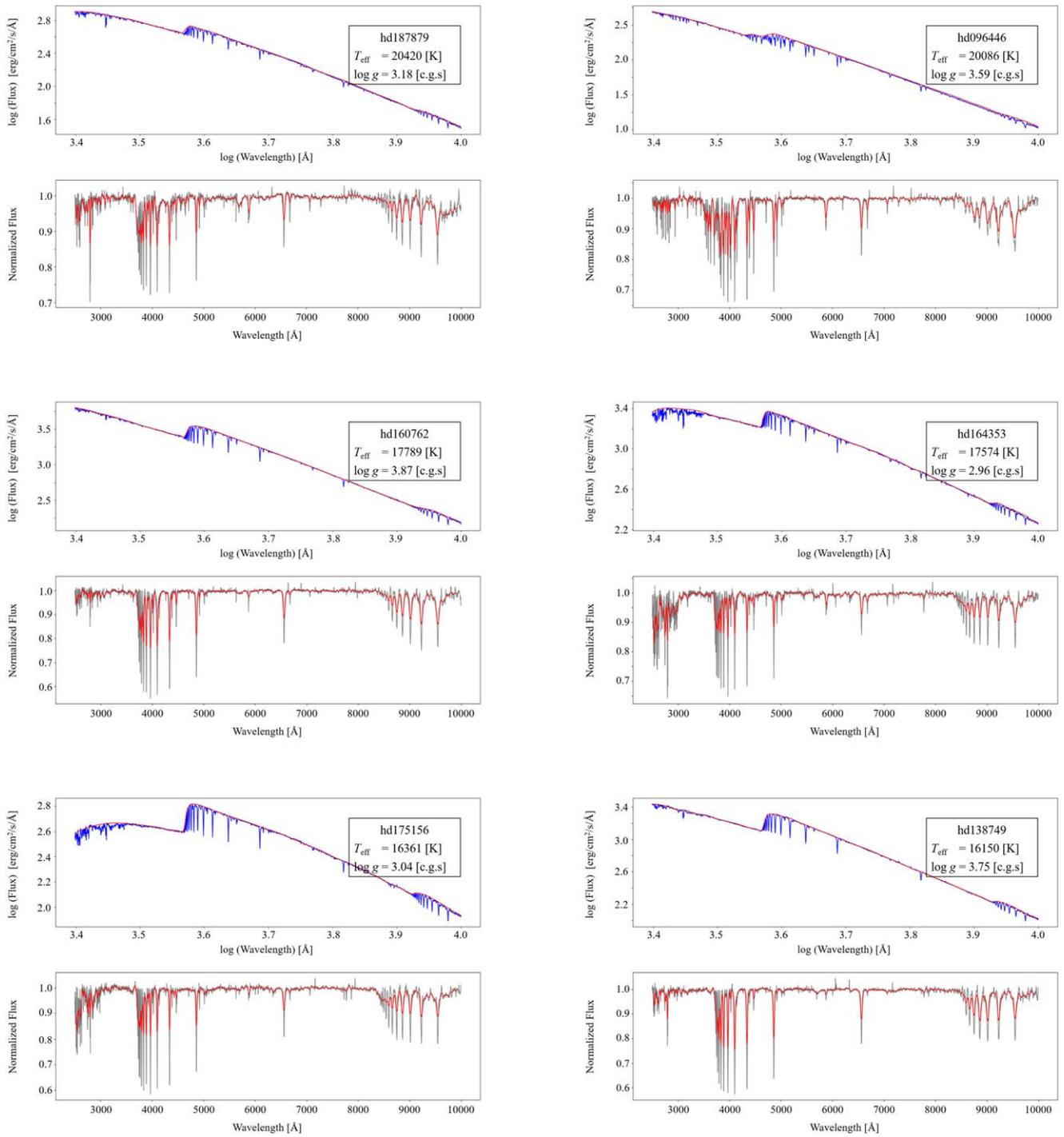

**Figure 14.** Similar to Figure 12.

## ORCID iDs

JiaRui Rao  https://orcid.org/0000-0003-3275-6622
JianPing Xiong  https://orcid.org/0000-0003-4829-6245
LuQian Wang  https://orcid.org/0000-0003-4511-6800
YanJun Guo  https://orcid.org/0000-0001-9989-9834
JiaJia Li  https://orcid.org/0009-0009-7824-5984
Chao Liu  https://orcid.org/0000-0002-1802-6917
ZhanWen Han  https://orcid.org/0000-0001-9204-7778
XueFei Chen  https://orcid.org/0000-0001-5284-8001